\documentclass[conference]{IEEEtran}
\IEEEoverridecommandlockouts
\usepackage{algorithmic}
\usepackage{amsmath,amssymb,amsfonts}
\usepackage{cite}
\usepackage{enumerate}
\usepackage{float}
\usepackage{graphicx}
\usepackage{multirow}
\usepackage[table,xcdraw]{xcolor}
\def\BibTeX{{\rm B\kern-.05em{\sc i\kern-.025em b}\kern-.08em
    T\kern-.1667em\lower.7ex\hbox{E}\kern-.125emX}}
    
\begin{document}

\title{Signal Propagation in RIS-aided 5G Systems\\
\thanks{The following work has been completed as part of project no.~$2021/43/\text{B}/\text{ST}7/01365$, funded by the National Science Centre (NSC) in Poland.}}

\author{\IEEEauthorblockN{Adam Samorzewski}
\IEEEauthorblockA{\textit{Institute of Radiocommunications} \\
\textit{Poznan University of Technology}\\
Poznan, Poland \\
adam.samorzewski@put.poznan.pl}
\and
\IEEEauthorblockN{Adrian Kliks}
\IEEEauthorblockA{\textit{Institute of Radiocommunications} \\
\textit{Poznan University of Technology}\\
Poznan, Poland \\
adrian.kliks@put.poznan.pl}
}

\maketitle

\begin{abstract}
In this paper, we conduct an in-depth analysis of radio signal propagation characteristics within the urban environment of Poznan (Poland). The study specifically addresses the deployment of a 5th generation (5G NR -- New Radio) Radio Access Network (RAN), which comprises 8 strategically positioned Base Stations (BSs). These base stations are configured with either Single Input Single Output (SISO) or Multiple Input Multiple Output (MIMO) antenna technologies, contingent upon the specific requirements of the network cells they serve. A key focus of our research is the integration of 15 reflecting arrays, known as Reconfigurable Intelligent Surfaces (RISs), which were installed throughout the study area. These RISs were deployed at various suspension heights to evaluate their impact on radio signal propagation and coverage. By exploring the influence of these RIS matrices, our research sheds light on their potential to significantly enhance signal quality, particularly in urban environments.\footnote{Copyright © 2024 IEEE. Personal use is permitted. For any other purposes, permission must be obtained from the IEEE by emailing pubs-permissions@ieee.org. This is the author’s version of an article that has been published in the proceedings of the 2024 20th International Conference on Wireless and Mobile Computing, Networking and Communications (WiMob) by the IEEE. Changes were made to this version by the publisher before publication, the final version of the record is available at: https://dx.doi.org/10.1109/WiMob61911.2024.10770363. To cite the paper use: A. Samorzewski, A.~Kliks, “Signal Propagation in RIS-Aided 5G Systems,” in: \textit{20th International Conference on Wireless and Mobile Computing, Networking and Communications (WiMob 2024)}, Paris, France, 2024, pp.~443--448, doi: 10.1109/WiMob61911.2024.10770363 or visit https://ieeexplore.ieee.org/document/10770363.}
\end{abstract}


\section{Introduction}
\label{section_introduction}
Research into current ($5$G) and future ($6$G) generations of mobile systems is introducing increasingly novel signal propagation solutions into the scientific literature. In recent years, Reconfigurable Intelligent Surfaces have emerged as a revolutionary technology promising to redefine wireless communication systems. Unlike traditional antenna technologies that rely on beamforming and power control at the transmitter, RIS introduces a~new paradigm by manipulating the propagation environment. These surfaces consist of large arrays of passive (or active) elements, such as metamaterials or simple reflectors, that can adaptively alter the electromagnetic properties of the environment to enhance signal transmission and reception. The concept of RIS draws inspiration from concepts in smart and reflective surfaces, integrating advanced signal processing and optimization techniques to dynamically control electromagnetic waves. By strategically adjusting the phase shifts and amplitude of reflected signals, RIS can focus and steer beams, compensate for channel impairments, and mitigate path losses across various propagation scenarios. Such a feature could enable Mobile Network Operators (MNOs) to manipulate radio signals to cover destination areas where reception devices have limited access to mobile services. Additionally, the use of RIS arrays can effectively reduce inter-signal interference and/or the exposure of users to electromagnetic fields (EMF). These arrays could be particularly useful in urban areas, where transmitted radio signals are often refracted and scattered by numerous obstacles (e.g., buildings). However, in rural areas, active RIS devices (those that amplify the redirected radio signal) could be deployed as relays, thereby reducing the number of base stations needed to cover the considered area \cite{Huang, DiRenzo, Tang, Liu, SamorzewskiJTIT2023, SamorzewskiKRiT2023, SamorzewskiSoftCOM2023, SamorzewskiGLOBECOM2023}.

This paper provides the fundamental principles behind RIS technology, discusses its potential applications in future wireless networks, and presents recent advancements and challenges in its implementation. The aim is to highlight how RIS can revolutionize communication systems by offering unprecedented flexibility, efficiency, and performance gains in real-world scenarios.

In this paper, the use of RIS matrices in an urban area is analyzed, examining their impact on the Path Loss (PL) characteristics of the propagated radio signal. The study also considers the effect of RIS suspension on the coverage of the studied area.

The article is divided into the following parts: Section~\ref{section_review} delivers an extensive literature review focusing on key positions of high interest; Section~\ref{section_scenario} describes the considered system scenario; Section~\ref{section_simulation} provides information on the research methodology and simulation configurations; Section~\ref{section_results} presents the research results in the form of graphs and tables with average values of observed system parameters; Section~\ref{section_conclusions} concludes the work by summarizing the findings based on the observed results.
\begin{figure*}[htb]
\centering
\includegraphics[width=\textwidth]{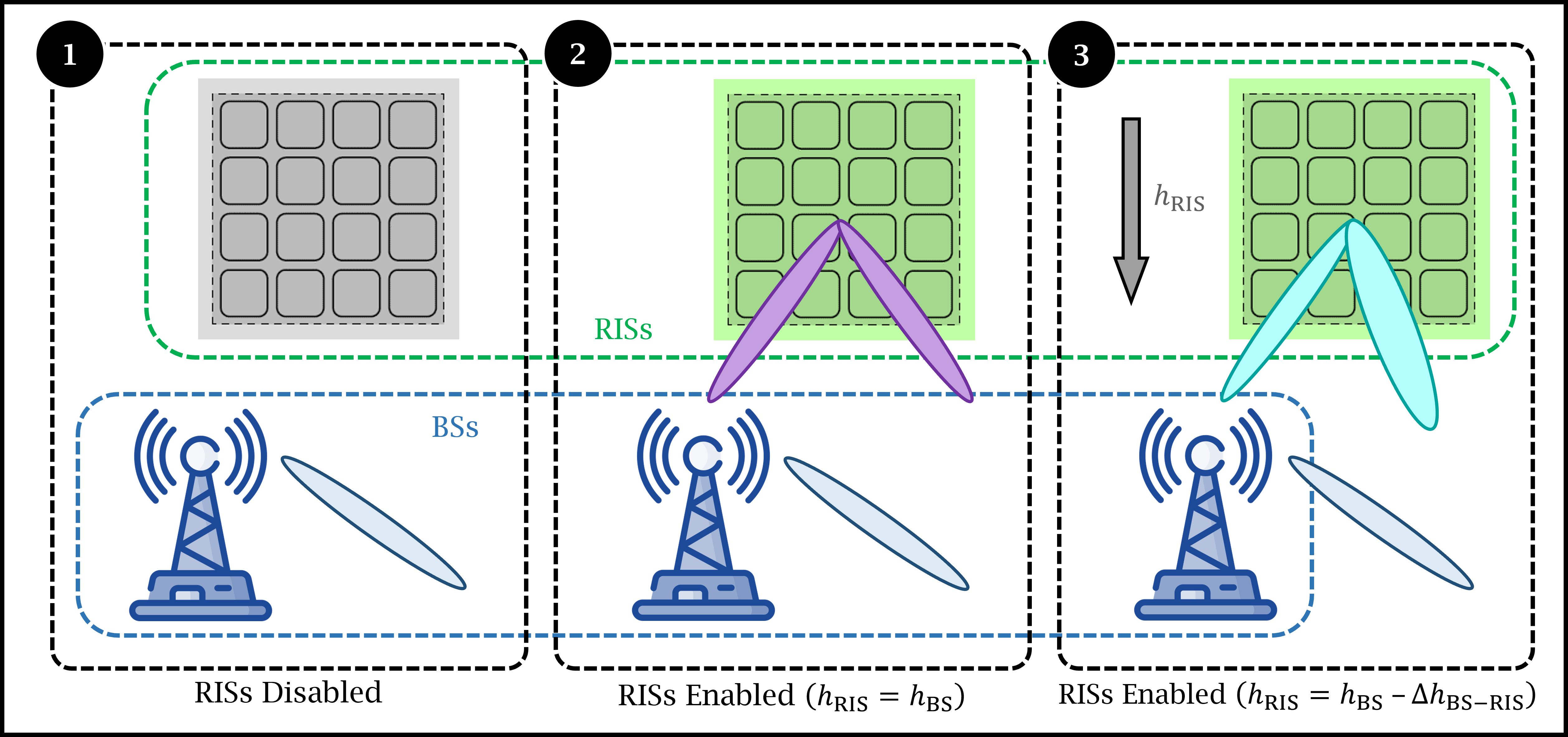}
\caption{Research methodology schema}
\label{figure_methodology_schema}
\end{figure*}

\section{Literature Review}
\label{section_review}
Nowadays, Reconfigurable Intelligent Surfaces have gained significant interest in scientific literature. The paper attached in \cite{Huang} investigates a downlink multiuser Multiple Input Single Output (MISO) system that employs RISs to maximize Energy Efficiency (EE). The authors present two computationally efficient algorithms to optimize base station transmit power allocation and RIS reflector values, enhancing the energy efficiency of wireless communication systems.

Next, the paper \cite{DiRenzo} explores the concept and applications of Reconfigurable Intelligent Surfaces in wireless communications. The work emphasizes the need to integrate C. E. Shannon's mathematical theory of communication with G.~Green’s and J.~C.~Maxwell’s theories of electromagnetism. Additionally, it offers practical guidelines for using physics-based models of metasurfaces in wireless networks.

Furthermore, in \cite{Tang} the authors have developed theoretical models for free-space path loss in RIS-assisted wireless communications, focusing on the electromagnetic and physics properties of RIS matrices. They proposed a general formula to determine the free-space path loss across various scenarios.

In addition, the work included in \cite{Liu} reviews recent advancements in RIS-enhanced wireless networks, focusing on operating principles, performance evaluation of multi-antenna systems, beamforming, resource allocation, and machine learning applications. It highlights the advantages and challenges of RISs, particularly in bridging complex physical models with communication models, and the difficulty of obtaining perfect Channel State Information (CSI). The integration of RISs with key 6G technologies like Non-Orthogonal Multiple Access (NOMA) and UAV-terrestrial (UAV \text{--} Unmanned Aerial Vehicle) networks is also discussed.

\section{Considered Scenario}
\label{section_scenario}
The considered network was deployed in the city of Poznan, specifically focusing on the Old Market Square and the surrounding streets \cite{SamorzewskiKRiT2023, SamorzewskiSoftCOM2023, SamorzewskiGLOBECOM2023}. The mobile system comprises $8$~base stations, each with $3$ cells transmitting radio signals at different frequencies: $800$, $2100$, or $3500$ MHz. Additionally, each of these cells employs antenna systems using either SISO ($800$ and $2100$ MHz) or MIMO ($3500$ MHz) techniques, characterized by different numbers of active elements: $1$ (SISO) and $64$~(MIMO) per transceiver \cite{Castellanos, SamorzewskiGLOBECOM2023}.

Furthermore, $15$ passive RIS arrays were strategically placed between buildings in the studied area, ensuring each array was within the range of at least one base station. These arrays are typically mounted on frames attached to the rooftops of buildings at the height of the base stations, although their elevation can be adjusted.

To design the mobile system, including determining the locations of base stations and buildings, as well as defining the dimensions of structures within the wireless system and the planned coverage area, real data from the BTSearch database \cite{NetworkData} and the Spatial Information System of the City of Poznan \cite{AreaData} were utilized.

\section{Simulation Configuration}
\label{section_simulation}
The research was conducted in a simulated form using dedicated software called Green Radio Access Network Design (GRAND) tool \cite{Castellanos, SamorzewskiGLOBECOM2023}.

The main objective of the study was to analyze the characteristics of radio signal path loss in the studied urban area. The evaluation of radio signal path loss values at specific Measurement Points (MPs) employed the following propagation models:
\begin{itemize}
\item $3$GPP TR $38.901$ UMa (Urban Macro) \text{--} signal propagation in Line-Of-Sight (LOS) and Non-Line-Of-Sight (NLOS) scenarios {\it directly} from the base station \cite{3GPP},
\item RIS-FFBC (RIS Far Field Beamforming Case) \text{--} signal propagation from the base station via reflection from Reconfigurable Intelligent Surfaces~\cite{Tang}.
\end{itemize}

Furthermore, the propagation conditions were arranged (simulated) for a system scenario where initially none of the available RIS units were used for indirect radio signal propagation, and then all of them were considered during the estimation of the path loss characteristics. Additionally, for the considered scenario, several variants of simulation configurations were tested, differing in the height at which RIS units reflecting the radio signal were mounted (the methodology has been shown in Fig.~\ref{figure_methodology_schema}). Each of these considered cases is outlined below $\left(h_\text{RIS}=h_\text{BS}-\Delta h_\text{BS--RIS}\right)$:
\begin{enumerate}[I.]
\item RISs mounted at height $h_\text{RIS}=h_\text{BS}\text{ }\left[\text{m}\right]$.
\item RISs mounted at height $h_\text{RIS}=h_\text{BS}-10\text{ }\left[\text{m}\right]$.
\item RISs mounted at height $h_\text{RIS}=h_\text{BS}-20\text{ }\left[\text{m}\right]$.
\item RISs mounted at height $h_\text{RIS}=h_\text{BS}-30\text{ }\left[\text{m}\right]$.
\end{enumerate}

The diverse set of test cases listed above will allow for an analysis of the impact of both the height of mounting RIS devices and the average difference in heights between the mounting of base station antenna arrays and RISs reflecting the radio beam from these access nodes (effectively within their range) on the propagation characteristics of the radio signal within the considered coverage area.

The configuration of simulation parameters for the considered mobile network is detailed in the tables presented below (i.e. Tab.~\ref{table_configuration_network_design} and \ref{table_configuration_ris_design}). The first one (Tab.~\ref{table_configuration_network_design}) includes the parameter values used to emulate the behavior of base stations in the simulation software. The parameters listed in this table provide information on the frequency band used, transceivers' configuration, and modeling of propagation conditions for each type of network cell. Meanwhile, Tab.~\ref{table_configuration_ris_design} comprises values for simulation parameters proposing a specifically defined type of Reconfigurable Intelligent Surface. Among these properties are characteristics such as the number and dimensions of reflecting elements, the height at which RIS units are mounted on buildings, and the propagation model used to estimate potential signal loss.

\section{Results}
\label{section_results}
\begin{figure}[htb]
\centering
\includegraphics[width=0.48\textwidth]{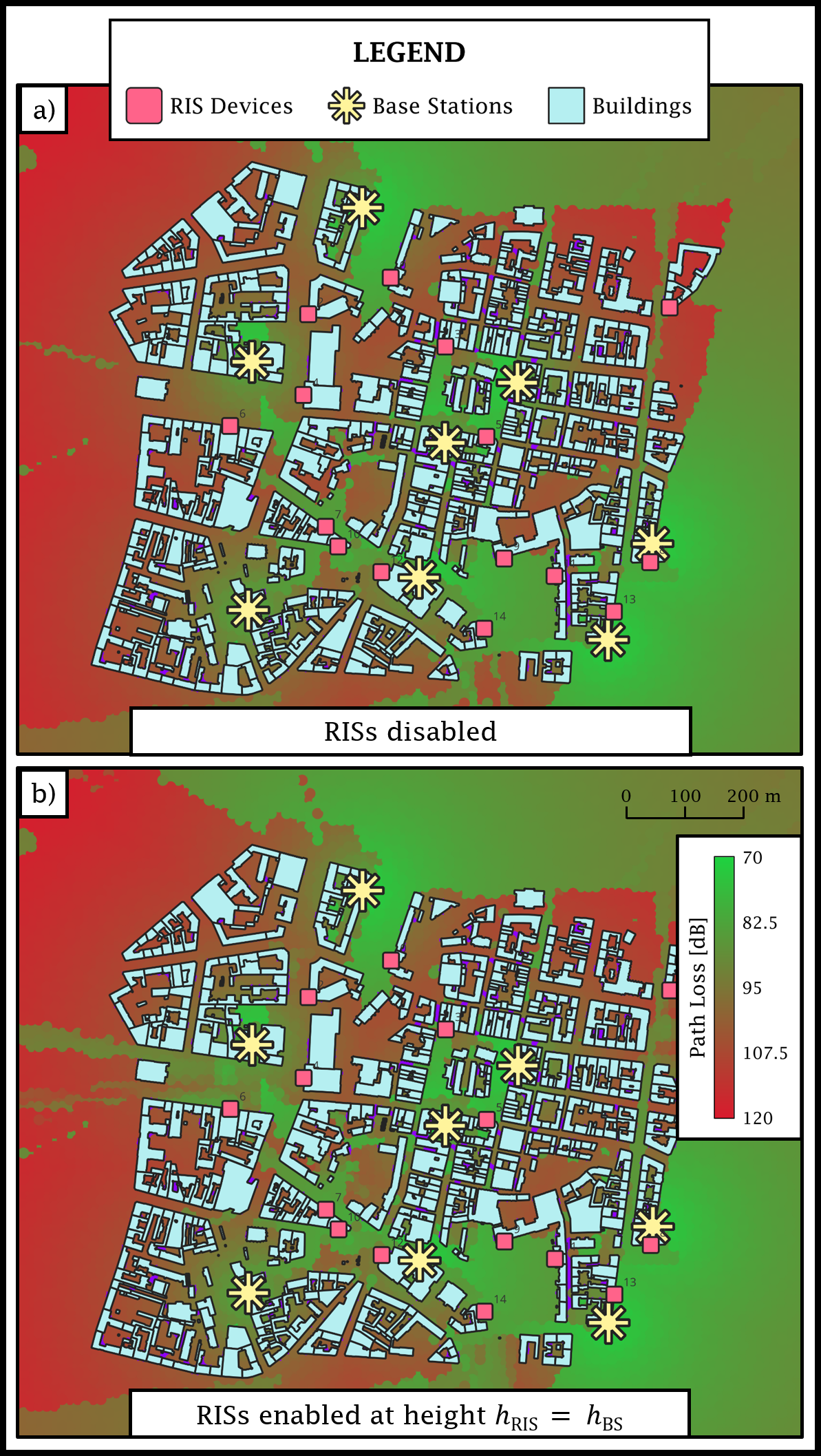}
\caption{Path Loss (PL) distribution for disabled and enabled RISs}
\label{figure_pl_dist_ris_no_ris}
\end{figure}

\begin{figure}[tbp]
\centering
\includegraphics[width=0.4825\textwidth]{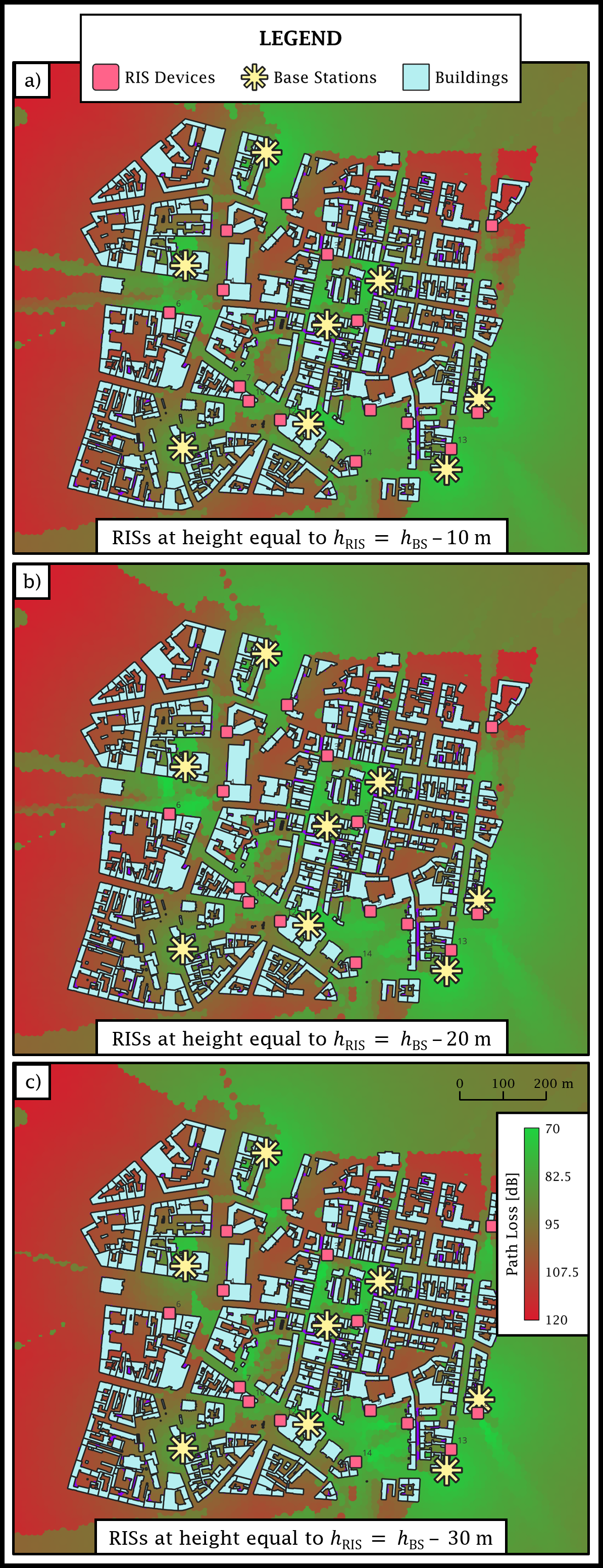}
\caption{PL dist. for diff. RIS heights $\left(h_\text{RIS}=h_\text{BS}-10/20/30\text{ m}\right)$}
\label{figure_pl_dist_ris_diff_h}
\end{figure}

This section presents the results obtained from conducting simulation studies. Fig.~\ref{figure_pl_dist_ris_no_ris}.a illustrates the distribution of minimum radio signal attenuation across the considered area when no reflecting arrays (i.e. RISs) are in use. Subsequently, Fig.~\ref{figure_pl_dist_ris_no_ris}.b and Fig.~\ref{figure_pl_dist_ris_diff_h}.a\text{--}c show the same results but with the inclusion of RIS for approximating the propagation characteristics, without and with adjustments in their hanging heights, respectively.

Comparing the maps in Fig.~\ref{figure_pl_dist_ris_no_ris} and \ref{figure_pl_dist_ris_diff_h}, it can be observed that introducing RISs has reduced signal attenuation in some regions of the studied area. Additionally, it is evident that the mounting height of intelligent surfaces significantly influences the presented radio signal propagation characteristics. Analyzing each case separately (Fig.~\ref{figure_pl_dist_ris_diff_h}.a\text{--}c), it is noticeable that decreasing the hanging height of RISs by the first $10$ meters and the subsequent $10$ meters ($20$ m relative to the default height \text{--} the height of BSs) reduces radio signal attenuation mostly in the southeast and western parts of the area. However, further lowering the attachment point of the reflecting arrays (by $30$~m relative to $h_\text{BS}$) unfortunately worsens this distribution in the western side of the area compared to Fig.~\ref{figure_pl_dist_ris_diff_h}.a\text{--}b. On the other hand, the signal propagation in different parts of the map (e.g., center \textbf{}{--} Old Market, northeast, and southeast) has been enhanced.
\begin{table}[htb]
\centering
\caption{Parameters configuration (BS)
\cite{Castellanos, SamorzewskiGLOBECOM2023, 3GPP, Bjornson, AreaData, NetworkData}
}
\label{table_configuration_network_design}
    \resizebox{0.4875\textwidth}{!}{\begin{tabular}{|l|c|c|c|cccc|}
    \arrayrulecolor[HTML]{002060}\hline
    \multirow{3}{*}{\cellcolor[HTML]{002060}}   & \multirow{3}{*}{\cellcolor[HTML]{002060}{\color[HTML]{FFFFFF}}} & \multirow{3}{*}{\cellcolor[HTML]{002060}{\color[HTML]{FFFFFF} }} & \multirow{3}{*}{\cellcolor[HTML]{002060}{\color[HTML]{FFFFFF} }} & \multicolumn{3}{c|}{\cellcolor[HTML]{002060}{\color[HTML]{FFFFFF} Value}}                                                                                                                                        \\ \cline{5-7} 
                        \cellcolor[HTML]{002060}{\color[HTML]{FFFFFF}} & \cellcolor[HTML]{002060}{\color[HTML]{FFFFFF} Parameter}                           &  \cellcolor[HTML]{002060}{\color[HTML]{FFFFFF} Sign}                     &  \cellcolor[HTML]{002060}{\color[HTML]{FFFFFF} Unit}                      & \multicolumn{3}{c|}{\cellcolor[HTML]{FFFFFF}{\color[HTML]{000000} \textit{Network Cell Type}}} \\ \cline{5-7}
                        \cellcolor[HTML]{002060}{\color[HTML]{FFFFFF}} & \cellcolor[HTML]{002060}{\color[HTML]{FFFFFF}}                           & \cellcolor[HTML]{002060}{\color[HTML]{FFFFFF}}                      & \cellcolor[HTML]{002060}{\color[HTML]{FFFFFF}}                      & \multicolumn{1}{c|}{$1\text{.}$}    & \multicolumn{1}{c|}{$2\text{.}$}   & \multicolumn{1}{c|}{$3\text{.}$} \\ \hline
    \multirow{1}{*}{\cellcolor[HTML]{89B0FF}{\color[HTML]{002060}}} 
                        \cellcolor[HTML]{89B0FF}{\color[HTML]{002060}} & \cellcolor[HTML]{E7E7E7}{\color[HTML]{000000} Quantity}                   & \cellcolor[HTML]{E7E7E7}{\color[HTML]{000000}$K_\text{BS}$}            & \cellcolor[HTML]{E7E7E7}{\color[HTML]{000000}\text{--}}                     &                                    \multicolumn{3}{c|}{\cellcolor[HTML]{E7E7E7}{\color[HTML]{000000}$8$}}      \\ \cline{2-7} 
                        \multirow{-2}{*}{\rotatebox[origin=c]{90}{\cellcolor[HTML]{89B0FF}{\color[HTML]{002060} Gen.}}} & Technology & \text{--} & \text{--} & \multicolumn{3}{c|}{$5\text{G}$} \\ 
                        \hline
    \multirow{1}{*}{\cellcolor[HTML]{89B0FF}{\color[HTML]{002060}}}
                        \cellcolor[HTML]{89B0FF}{\color[HTML]{002060}} & \cellcolor[HTML]{E7E7E7}{\color[HTML]{000000} Frequency}                  & \cellcolor[HTML]{E7E7E7}{\color[HTML]{000000} $f$}            & \cellcolor[HTML]{E7E7E7}{\color[HTML]{000000}{$\left[\text{MHz}\right]$}}             & \multicolumn{1}{c|}{\cellcolor[HTML]{E7E7E7}{\color[HTML]{000000} $\quad800\quad$}}  & \multicolumn{1}{c|}{\cellcolor[HTML]{E7E7E7}{\color[HTML]{000000}$\quad2100\quad$}} & \multicolumn{1}{c|}{\cellcolor[HTML]{E7E7E7}{\color[HTML]{000000}$\quad3500\quad$}} \\ \cline{2-7} 
                        \cellcolor[HTML]{89B0FF}{\color[HTML]{002060}} & Channel Bandwidth          & $B_\text{w}$            & {$\left[\text{MHz}\right]$}      & \multicolumn{1}{c|}{$80$}   & \multicolumn{1}{c|}{$120$}  & \multicolumn{1}{c|}{$120$} \\ \cline{2-7} 
                        \cellcolor[HTML]{89B0FF}{\color[HTML]{002060}} & \cellcolor[HTML]{E7E7E7}{\color[HTML]{000000} Used Subcarriers}           & \cellcolor[HTML]{E7E7E7}{\color[HTML]{000000} $N_\text{SC,u}$}            & \cellcolor[HTML]{E7E7E7}{\color[HTML]{000000} \text{--}}                     & \multicolumn{3}{c|}{\cellcolor[HTML]{E7E7E7}{\color[HTML]{000000} $320$}} \\ \cline{2-7} 
                        \cellcolor[HTML]{89B0FF}{\color[HTML]{002060}} & Total Subcarriers          & $N_\text{SC,t}$            & \text{--}                     & \multicolumn{3}{c|}{$512$} \\ \cline{2-7} 
                        \cellcolor[HTML]{89B0FF}{\color[HTML]{002060}} & \cellcolor[HTML]{E7E7E7}{\color[HTML]{000000} Sampling Factor}            & \cellcolor[HTML]{E7E7E7}{\color[HTML]{000000} SF}                    & \cellcolor[HTML]{E7E7E7}{\color[HTML]{000000} \text{--}}                     & \multicolumn{3}{c|}{\cellcolor[HTML]{E7E7E7}{\color[HTML]{000000} $1.536$}} \\ \cline{2-7} 
                        \cellcolor[HTML]{89B0FF}{\color[HTML]{002060}} & Pilot Reuse Factor         & RF                    & \text{--}                     & \multicolumn{3}{c|}{$1$} \\ \cline{2-7} 
                        \cellcolor[HTML]{89B0FF}{\color[HTML]{002060}} & \cellcolor[HTML]{E7E7E7}{\color[HTML]{000000} Coherence Time}             & \cellcolor[HTML]{E7E7E7}{\color[HTML]{000000} $t_\text{c}$}            & \cellcolor[HTML]{E7E7E7}{\color[HTML]{000000}{$\left[\text{ms}\right]$}}              & \multicolumn{3}{c|}{\cellcolor[HTML]{E7E7E7}{\color[HTML]{000000}$50$}} \\ \cline{2-7} 
                        \cellcolor[HTML]{89B0FF}{\color[HTML]{002060}} & Coherence Bandwidth        & $B_\text{c}$            & {$\left[\text{MHz}\right]$}             & \multicolumn{3}{c|}{$1$} \\ \cline{2-7}
                        \multirow{-9}{*}{\rotatebox[origin=c]{90}{\cellcolor[HTML]{89B0FF}{\color[HTML]{002060} Band}}} & \cellcolor[HTML]{E7E7E7}{\color[HTML]{000000} Spatial Duty Cycle}         & \cellcolor[HTML]{E7E7E7}{\color[HTML]{000000} $S$}            & \cellcolor[HTML]{E7E7E7}{\color[HTML]{000000}{$\left[\text{\%}\right]$}}              & \multicolumn{1}{c|}{\cellcolor[HTML]{E7E7E7}{\color[HTML]{000000} $0$}}    & \multicolumn{1}{c|}{\cellcolor[HTML]{E7E7E7}{\color[HTML]{000000} $0$}}    & \multicolumn{1}{c|}{\cellcolor[HTML]{E7E7E7}{\color[HTML]{000000} $25$}} \\ \hline
    \multirow{1}{*}{\cellcolor[HTML]{89B0FF}{\color[HTML]{002060}}}  & Antenna Height             & $h_\text{BS}$            & {$\left[\text{m}\right]$}               & \multicolumn{3}{c|}{$42.5$} \\ \cline{2-7} 
                        \cellcolor[HTML]{89B0FF}{\color[HTML]{002060}} & \cellcolor[HTML]{E7E7E7}{\color[HTML]{000000} Antenna Elements}           & \cellcolor[HTML]{E7E7E7}{\color[HTML]{000000} $M_\text{BS}$}            & \cellcolor[HTML]{E7E7E7}{\color[HTML]{000000} \text{--}}                     & \multicolumn{1}{c|}{\cellcolor[HTML]{E7E7E7}{\color[HTML]{000000} $1$}}    & \multicolumn{1}{c|}{\cellcolor[HTML]{E7E7E7}{\color[HTML]{000000} $1$}}    & \multicolumn{1}{c|}{\cellcolor[HTML]{E7E7E7}{\color[HTML]{000000} $64$}} \\ \cline{2-7} 
                        \cellcolor[HTML]{89B0FF}{\color[HTML]{002060}} & Antenna Gain               & $G_\text{a}$            & {$\left[\text{dBi}\right]$}             & \multicolumn{1}{c|}{$16$}   & \multicolumn{1}{c|}{$18$}   & \multicolumn{1}{c|}{$24$} \\ \cline{2-7} 
                        \cellcolor[HTML]{89B0FF}{\color[HTML]{002060}} & \cellcolor[HTML]{E7E7E7}{\color[HTML]{000000} Feeder Loss}        & \cellcolor[HTML]{E7E7E7}{\color[HTML]{000000} $L_\text{f}$}            & \cellcolor[HTML]{E7E7E7}{\color[HTML]{000000}{$\left[\text{dBi}\right]$}}             & \multicolumn{1}{c|}{\cellcolor[HTML]{E7E7E7}{\color[HTML]{000000} $2$}}    & \multicolumn{1}{c|}{\cellcolor[HTML]{E7E7E7}{\color[HTML]{000000} $2$}}    & \multicolumn{1}{c|}{\cellcolor[HTML]{E7E7E7}{\color[HTML]{000000} $3$}} \\ \cline{2-7} 
                        \cellcolor[HTML]{89B0FF}{\color[HTML]{002060}} & Max. Transmit Power        & $P_\text{TX, max}$            & {$\left[\text{dBm}\right]$}             & \multicolumn{1}{c|}{$46$}   & \multicolumn{1}{c|}{$49$}   & \multicolumn{1}{c|}{$53$} \\ \cline{2-7} 
                        \multirow{-6}{*}{\rotatebox[origin=c]{90}{\cellcolor[HTML]{89B0FF}{\color[HTML]{002060} Transceivers}}} & \cellcolor[HTML]{E7E7E7}{\color[HTML]{000000} Noise Factor}               & \cellcolor[HTML]{E7E7E7}{\color[HTML]{000000} NF}                    & \cellcolor[HTML]{E7E7E7}{\color[HTML]{000000}{$\left[\text{dB}\right]$}}              & \multicolumn{1}{c|}{\cellcolor[HTML]{E7E7E7}{\color[HTML]{000000} $8$}}    & \multicolumn{1}{c|}{\cellcolor[HTML]{E7E7E7}{\color[HTML]{000000} $8$}}    & \multicolumn{1}{c|}{\cellcolor[HTML]{E7E7E7}{\color[HTML]{000000} $7$}} \\ \hline
    \multirow{1}{*}{\cellcolor[HTML]{89B0FF}{\color[HTML]{002060}}}  & Path Loss Model            & \text{--}                     & \text{--}                     & \multicolumn{3}{c|}{$3$GPP TR $38.901$ UMa \cite{3GPP}} \\ \cline{2-7} 
                        \cellcolor[HTML]{89B0FF}{\color[HTML]{002060}} & \cellcolor[HTML]{E7E7E7}{\color[HTML]{000000} Interference Margin}        & \cellcolor[HTML]{E7E7E7}{\color[HTML]{000000} IM}                    & \cellcolor[HTML]{E7E7E7}{\color[HTML]{000000}{$\left[\text{dB}\right]$}}              & \multicolumn{3}{c|}{\cellcolor[HTML]{E7E7E7}{\color[HTML]{000000} $2$}}  \\ \cline{2-7} 
                        \cellcolor[HTML]{89B0FF}{\color[HTML]{002060}} & Doppler Margin             & DM                    & {$\left[\text{dB}\right]$}              & \multicolumn{3}{c|}{$3$} \\ \cline{2-7} 
                        \cellcolor[HTML]{89B0FF}{\color[HTML]{002060}} & \cellcolor[HTML]{E7E7E7}{\color[HTML]{000000} Fade Margin}                & \cellcolor[HTML]{E7E7E7}{\color[HTML]{000000} FM}                    & \cellcolor[HTML]{E7E7E7}{\color[HTML]{000000}{$\left[\text{dB}\right]$}}              & \multicolumn{3}{c|}{\cellcolor[HTML]{E7E7E7}{\color[HTML]{000000} $10$}} \\ \cline{2-7} 
                        \cellcolor[HTML]{89B0FF}{\color[HTML]{002060}} & Shadow Margin              & SM                    & {$\left[\text{dB}\right]$}              & \multicolumn{1}{c|}{$12.8$} & \multicolumn{1}{c|}{$15.2$} & \multicolumn{1}{c|}{$10$} \\ \cline{2-7} 
                       \multirow{-6}{*}{\rotatebox[origin=c]{90}{\cellcolor[HTML]{89B0FF}{\color[HTML]{002060} Propagation}}} & \cellcolor[HTML]{E7E7E7}{\color[HTML]{000000} Implementation Loss}         & \cellcolor[HTML]{E7E7E7}{\color[HTML]{000000} IL}            & \cellcolor[HTML]{E7E7E7}{\color[HTML]{000000}{$\left[\text{dB}\right]$}}              & \multicolumn{1}{c|}{\cellcolor[HTML]{E7E7E7}{\color[HTML]{000000} $0$}}    & \multicolumn{1}{c|}{\cellcolor[HTML]{E7E7E7}{\color[HTML]{000000} $0$}}    & \multicolumn{1}{c|}{\cellcolor[HTML]{E7E7E7}{\color[HTML]{000000} $3$}} \\ \hline
    \end{tabular}}
\end{table}
\begin{table}[htb]
\centering
\caption{Parameters configuration (RIS)
\cite{Tang}
}
\label{table_configuration_ris_design}
    \resizebox{0.3875\textwidth}{!}{\begin{tabular}{|l|c|c|c|c|}
    \arrayrulecolor[HTML]{002060}\hline
                        \cellcolor[HTML]{002060}{\color[HTML]{FFFFFF}} & \cellcolor[HTML]{002060}{\color[HTML]{FFFFFF} Parameter}                           &  \cellcolor[HTML]{002060}{\color[HTML]{FFFFFF} Sign}                     &  \cellcolor[HTML]{002060}{\color[HTML]{FFFFFF} Unit}                      & \cellcolor[HTML]{002060}{\color[HTML]{FFFFFF} Value} \\ \hline
                        \multirow{1}{*}{\cellcolor[HTML]{89B0FF}{\color[HTML]{002060}}} 
                        \cellcolor[HTML]{89B0FF}{\color[HTML]{002060}} & \cellcolor[HTML]{E7E7E7}{\color[HTML]{000000} Quantity}                   & \cellcolor[HTML]{E7E7E7}{\color[HTML]{000000} $K_\text{RIS}$}            & \cellcolor[HTML]{E7E7E7}{\color[HTML]{000000}\text{--}}                     &                                    \multicolumn{1}{c|}{\cellcolor[HTML]{E7E7E7}{\color[HTML]{000000}$15$}}      \\ \cline{2-5} 
                        \multirow{-6}{*}{\rotatebox[origin=c]{90}{\cellcolor[HTML]{89B0FF}{\color[HTML]{002060}}}} & {\color[HTML]{000000} Column Elements}         & {\color[HTML]{000000} $M_\text{RIS}$}            & {\color[HTML]{000000}{\text{--}}}              & \multicolumn{1}{c|}{{\color[HTML]{000000} $102$}} \\ \cline{2-5} 
                        \cellcolor[HTML]{89B0FF}{\color[HTML]{002060}} & \cellcolor[HTML]{E7E7E7}{\color[HTML]{000000} Row Elements}        & \cellcolor[HTML]{E7E7E7}{\color[HTML]{000000} $N_\text{RIS}$}                    & \cellcolor[HTML]{E7E7E7}{\color[HTML]{000000}{\text{--}}}              & \multicolumn{1}{c|}{\cellcolor[HTML]{E7E7E7}{\color[HTML]{000000} $100$}}  \\ \cline{2-5} 
                        \cellcolor[HTML]{89B0FF}{\color[HTML]{002060}} & Column Width             & $d_\text{m}$                    & {$\left[\text{m}\right]$}              & \multicolumn{1}{c|}{$0.01$} \\ \cline{2-5} 
                        \cellcolor[HTML]{89B0FF}{\color[HTML]{002060}} & \cellcolor[HTML]{E7E7E7}{\color[HTML]{000000} Row Width}                & \cellcolor[HTML]{E7E7E7}{\color[HTML]{000000} $d_\text{n}$}                    & \cellcolor[HTML]{E7E7E7}{\color[HTML]{000000}{$\left[\text{m}\right]$}}              & \multicolumn{1}{c|}{\cellcolor[HTML]{E7E7E7}{\color[HTML]{000000} $0.01$}} \\ \cline{2-5} 
                        \cellcolor[HTML]{89B0FF}{\color[HTML]{002060}} & Reflected Signal Amp. Factor         & $A$                    & {\text{--}}              & \multicolumn{1}{c|}{$0.9$} \\ \cline{2-5}
                        \cellcolor[HTML]{89B0FF}{\color[HTML]{002060}} & \cellcolor[HTML]{E7E7E7}{\color[HTML]{000000} Suspension Height} & \cellcolor[HTML]{E7E7E7}{\color[HTML]{000000} $h_\text{RIS}$}                    & \cellcolor[HTML]{E7E7E7}{\color[HTML]{000000} $\left[\text{m}\right]$}              & \multicolumn{1}{c|}{\cellcolor[HTML]{E7E7E7}{\color[HTML]{000000} $\left(12.5, 42.5\right)$}} \\ \cline{2-5}  
                        \multirow{1}{*}{\cellcolor[HTML]{89B0FF}{\color[HTML]{002060}}}  & {\color[HTML]{000000} Path Loss Model} & {\color[HTML]{000000} \text{--}}                     & {\color[HTML]{000000} \text{--}}     & \multicolumn{1}{c|}{{\color[HTML]{000000} RIS-FFBC \cite{Tang}}}\\ \hline
    \end{tabular}}
\end{table}

The above observations are summarized in Tab.~\ref{table_results}. The values of parameters in this table were obtained using the following formulas, marked as Eq.~(\ref{equation_delta_h_bs_ris}) and (\ref{equation_delta_pl_avg}).

The very first parameter described in Tab.~\ref{table_results} is the height difference between base stations and RIS arrays $\left(\Delta h_\text{BS--RIS}\right)$, which directed the radio signal from these stations to various points on the map of the considered area. This parameter is described by the equation:
\begin{equation}
\label{equation_delta_h_bs_ris}
    \Delta h_\text{BS--RIS}=h_{\text{BS}}-h_{\text{RIS}}^\text{current},
\end{equation}
where $h_{\text{BS}}$ is the global height of the antennas of the network cells, and $h_{\text{RIS}}^\text{current}$ is the current height of the RIS devices (both in m).

The next and more important evaluation parameter, the average gain associated with the reduction in radio signal path loss $\left(\Delta \overline{\text{PL}}\right)$, was implemented in the simulation environment as follows:
\begin{equation}
\label{equation_delta_pl_avg}
    \Delta \overline{\text{PL}}=\Bigg[1-\frac{\sum^{Y-1}_{y=0}\sum^{X-1}_{x=0}\text{PL}_\text{RIS}^\text{min}\left(x,y\right)}{\sum^{Y-1}_{y=0}\sum^{X-1}_{x=0}\text{PL}_\text{NRIS}^\text{min}\left(x,y\right)}\Bigg] \cdot 100\text{ }\%,
\end{equation}
where $x$ $\left(x=0,\dots,X-1\right)$ and $y$ $\left(y=0,\dots,Y-1\right)$ are coordinates of a given point within the studied urban area. $\text{PL}_\text{RIS}^\text{min}\left(x,y\right)$ and $\text{PL}_\text{NRIS}^\text{min}\left(x,y\right)$ are values of radio signal propagation losses (in dB) at the point $P\left(x,y\right)$ when the influence of RISs on the signal transmission is considered and when this impact is omitted, respectively. These parameters are defined within the simulation software as indicated by the following formulas:
\begin{align}
\label{equation_pl_ris_min}
    &\text{PL}_\text{RIS}^\text{min}\left(x,y\right)=\\ &\min_{i,j}\Bigl\{\text{PL}_{\text{LOS},i}\left(x,y\right);\text{}\text{PL}_{\text{NLOS},i}\left(x,y\right);\text{}\text{PL}_{\text{RIS},i,j}\left(x,y\right)\Bigr\}, \nonumber
\end{align}
\begin{equation}
\label{equation_pl_no_ris_min}
    \text{PL}_\text{NRIS}^\text{min}\left(x,y\right)=\min_{i}\Bigl\{\text{PL}_{\text{LOS},i}\left(x,y\right);\text{}\text{PL}_{\text{NLOS},i}\left(x,y\right)\Bigr\},
\end{equation}
\begin{figure*}[tbp]
\centering
\includegraphics[width=\textwidth]{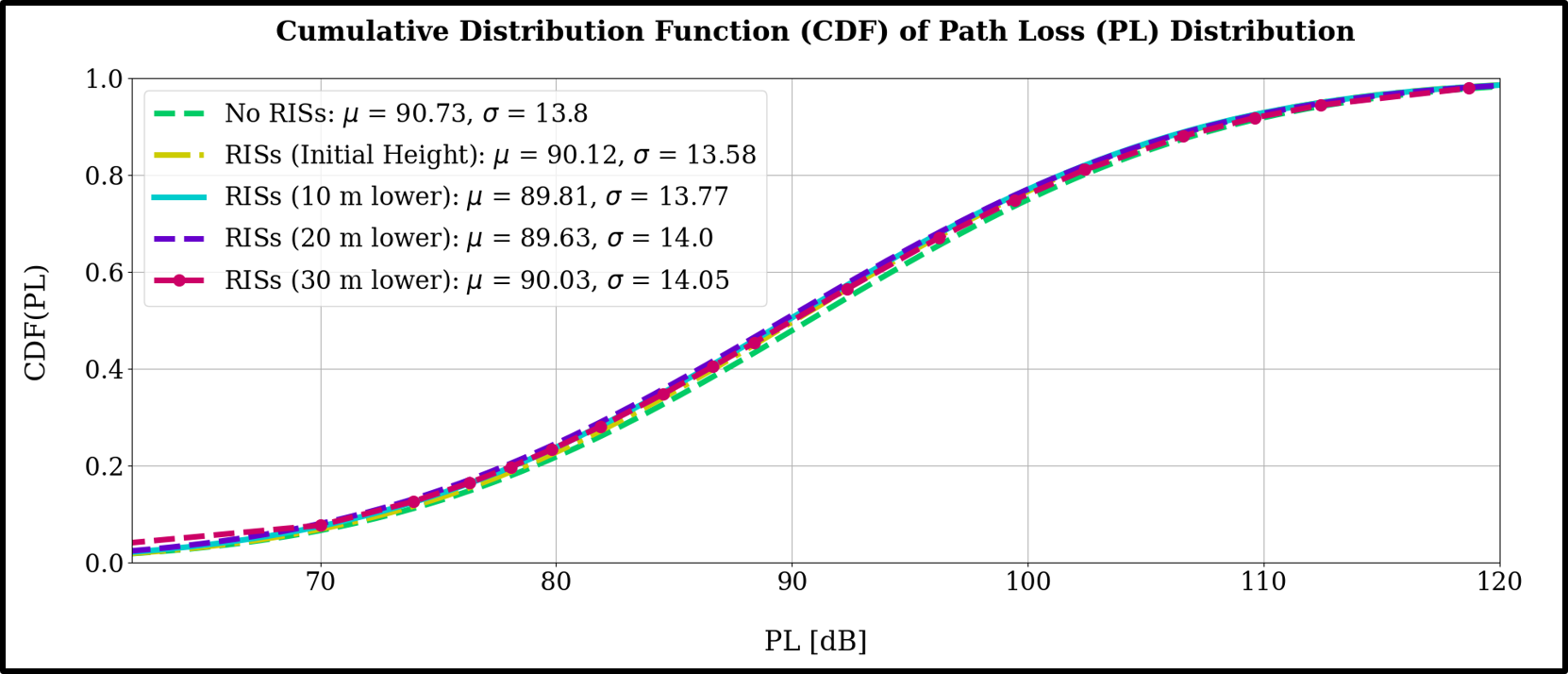}
\caption{CDF plots of PL distribution with RISs disabled and enabled at various suspension heights}
\label{figure_cdf_pl}
\end{figure*}
where $\text{PL}_{\text{LOS},i}\left(x,y\right)$, $\text{PL}_{\text{NLOS},i}\left(x,y\right)$, and $\text{PL}_{\text{RIS},i,j}\left(x,y\right)$ are the values for radio signal attenuation measured at point $P\left(x,y\right)$ from the $i$-th $\left(i=0,\dots,K_\text{BS}-1\right)$ network cell \textit{directly} \text{--} via LOS \text{--} $\left(\text{PL}_{\text{LOS},i}\right)$, \textit{indirectly} \text{--} via NLOS, i.e. after its power has been attenuated by physical objects (e.g., buildings) or atmospheric phenomena \text{--} $\left(\text{PL}_{\text{NLOS},i}\right)$, and after reflection of the radio beam from the $j$-th $\left(j=0,\dots,K_\text{RIS}-1\right)$ RIS device $\left(\text{PL}_{\text{RIS},i,j}\right)$. In both cases \text{--} Eq.~(\ref{equation_pl_ris_min}) and (\ref{equation_pl_no_ris_min}) \text{--} the more favorable (minimal) path loss value is always chosen. The parameters of $K_\text{BS}$ and $K_\text{RIS}$ are the numbers of network cells and RIS matrices within the system, respectively.

Tab.~\ref{table_results} summarizes the impact of varying the height difference $\left(\Delta h_\text{BS--RIS}\right)$ between RISs and BSs on the path loss gain $\left(\Delta\overline{\text{PL}}\right)$ compared to scenarios without RISs enabled. The heights considered are $0$ m, $10$ m, $20$ m, and $30$ m. The results show that as the height difference increases, there is an initial rise in the gain related to radio signal attenuation reduction, with the scenario at $20$ m $\left(\Delta \overline{\text{PL}}=1.21\text{ }\%\right)$ demonstrating the highest gain. However, at $30$ m $\left(\Delta \overline{\text{PL}}=0.77\text{ }\%\right)$, a~slight decrease is observed compared to the $20$ m scenario, suggesting a~potential optimal height range for maximizing path loss reduction.

Referring directly to Tab.~\ref{table_results}, it can be observed that after surpassing the threshold where the height difference between base stations and RISs $\left(\Delta h_\text{BS--RIS}\right)$ increases from $20$ to $30$ meters, the gain associated with the average received radio signal power in the considered area $\left(\Delta\overline{\text{PL}}\right)$ begins to decline by approximately $0.44$~pp (percentage points) compared to the most favorable scenario (RISs positioned 20 meters lower than the BSs). This trend highlights the significance of precise RIS positioning, indicating that their ability to effectively manipulate electromagnetic waves and enhance signal propagation is highly dependent on their relative height to the BSs.

Such insights are crucial for optimizing RIS deployment strategies in future wireless communication systems, as they underscore the importance of carefully balancing the enhancement of signal coverage with the minimization of path loss. Understanding the nuanced relationship between RIS height and path loss gain can guide the design of more efficient network configurations, particularly in urban environments where obstacles and varied building heights present unique challenges to signal propagation. Consequently, these findings contribute to the broader effort to refine RIS technology and its application in improving the performance and reliability of next-generation wireless networks.
\begin{table}[htb]
\centering
\caption{Impact of the RIS height on the average Path Loss}
\label{table_results}
\resizebox{0.4875\textwidth}{!}{
\begin{tabular}{|c|c|c|c|c|}
\arrayrulecolor[HTML]{002060}
\hline
{\cellcolor[HTML]{002060}{\color[HTML]{FFFFFF} Parameter}} & \multicolumn{4}{c|}{\cellcolor[HTML]{002060}{\color[HTML]{FFFFFF} Value}} \\ \cline{2-5} 
\cellcolor[HTML]{FFFFFF}{\color[HTML]{000000} $\Delta h_\text{BS--RIS}$} & \cellcolor[HTML]{FFFFFF}{\color[HTML]{000000} $0\text{ m}$} & \cellcolor[HTML]{FFFFFF}{\color[HTML]{000000} $10\text{ m}$} & \cellcolor[HTML]{FFFFFF}{\color[HTML]{000000} $20\text{ m}$} & \cellcolor[HTML]{FFFFFF}{\color[HTML]{000000} $30\text{ m}$} \\ \hline
\cellcolor[HTML]{89B0FF}{\color[HTML]{002060} $\Delta \overline{\text{PL}}$} & \cellcolor[HTML]{89B0FF}{\color[HTML]{002060} $0.67\text{ }\%$} & \cellcolor[HTML]{89B0FF}{\color[HTML]{002060} $1.01\text{ }\%$} & \cellcolor[HTML]{89B0FF}{\color[HTML]{002060} $1.21\text{ }\%$} & \cellcolor[HTML]{89B0FF}{\color[HTML]{002060} $0.77\text{ }\%$} \\ \hline
\end{tabular}}
\end{table}

Furthermore, in Fig.~\ref{figure_cdf_pl}, the Cumulative Distribution Function (CDF) for path loss distribution within the examined area is presented for different RIS heights. Comparing the CDF plots of path loss distributions across these scenarios in Poznan reveals that deploying RIS at lower heights generally (but slightly) reduces path loss, thereby improving signal strength and coverage. This enhancement is particularly noticeable in densely built-up areas where lower RIS heights help mitigate obstructions and provide more direct signal paths. However, the optimal height for RIS deployment depends on balancing several factors, including deployment costs, installation complexity, and the specific urban propagation characteristics of the area. While the plots do not show dramatic differences between the various heights, the trend observed aligns with the findings summarized in Tab.~\ref{table_results}, confirming that decreasing the height of the RIS matrices results in improved path loss characteristics for the considered mobile network scenario. This observation highlights the importance of strategic RIS placement in optimizing network performance, emphasizing that even subtle adjustments in deployment height can lead to measurable benefits in signal quality and coverage, particularly in challenging urban environments.

Next to the line markers denoting all simulation configurations -- including No RISs, RISs (Initial Height), RISs ($10$~m lower), RISs ($20$ m lower), and RISs ($30$ m lower) -- the parameters of each CDF function, specifically the mean value $\left(\mu\right)$ and standard deviation $\left(\sigma\right)$, have been explicitly detailed. These statistical parameters provide critical insights into the overall behavior of path loss across different scenarios. The mean value offers an indication of the average path loss experienced within the examined area, allowing for a~comparative assessment of how RIS deployment at various heights influences signal attenuation. Meanwhile, the standard deviation reflects the variability in path loss, shedding light on the consistency of signal strength across different locations within the urban environment.

By analyzing these parameters, we can better understand the extent to which lowering RIS height impacts not only the average signal strength but also the uniformity of coverage. For instance, a lower standard deviation in scenarios where RISs are deployed at reduced heights might suggest that these configurations not only enhance the mean signal strength but also contribute to more stable and reliable signal propagation throughout the area. This comprehensive statistical analysis, which is intended to be extensively explored in future research, plays a crucial role in determining the optimal RIS deployment strategy, ensuring that both overall performance and coverage consistency are maximized in real-world network scenarios.

\section{Conclusions}
\label{section_conclusions}
Deploying Reconfigurable Intelligent Surfaces in urban areas like Poznan enhances signal quality by reducing path loss and improving coverage, which is essential in densely populated regions with high demand for reliable connectivity. These enhancements are vital for supporting the growing number of devices and services that depend on stable and robust wireless networks in such areas. However, challenges arise from the complexity of integrating RISs into existing infrastructure, often requiring substantial modifications to accommodate the new technology. Moreover, obtaining accurate CSI in dynamic urban environments, where signal conditions can rapidly change due to factors like moving vehicles and varying user densities, presents a significant technical challenge.

The study demonstrates that RISs can slightly improve signal quality, particularly when positioned at optimal heights that maximize their ability to reflect and direct signals effectively. However, the benefits diminish beyond certain thresholds, indicating that there is a limit to how much height adjustment can enhance performance. While adjusting RIS height improves signal transmission by optimizing LOS and reducing the impact of obstacles, careful planning and precise calculations are necessary to maximize these gains and prevent diminishing returns.

Future research will concentrate on optimizing RIS deployment by evaluating their impact on service quality, including throughput, latency, and user experience, as well as their contribution to energy efficiency. This research will also explore how RISs can be integrated with energy-efficient technologies and Renewable Energy Sources (RES) \text{--} e.g., Wind Turbines (WTs) and Photovoltaic (PV) panels \text{--} to build more sustainable networks. Additionally, simulations will be validated against real-world data to ensure that theoretical models accurately reflect practical outcomes, ultimately guiding the development of more efficient and resilient networks to meet the demands of future urban environments. The research will further explore the potential of RISs to support emerging technologies such as 5G and beyond, smart cities, and the Internet of Things (IoT), emphasizing their crucial role in the future of urban connectivity.

\section*{Acknowledgements}
\label{section_ackonwledgements}
The authors express their gratitude to Prof. Margot Deruyck from Ghent University (IMEC) in Belgium for her invaluable support of this work through the provision of the GRAND software, which significantly enhanced the research outcomes.

Adrian Kliks is also affiliated with the Luleå University of Technology in Sweden, where he contributes to research and academic initiatives.


\begin{thebibliography}{00}
\bibitem[1]{Huang} C. Huang, A. Zappone, G.C. Alexandropoulos, M.~Debbah, and C. Yuen, "Reconfigurable Intelligent Surfaces for Energy Efficiency in Wireless Communication," {\it IEEE Transactions on Wireless Communications}, vol. $18$, no. $8$, pp. $4157$\textbf{–-}$4170$, $2019$. DOI: $10$.$1109$/TWC.$2019$.$2922609$.

\bibitem[2]{DiRenzo} M. Di Renzo et al., "Smart Radio Environments Empowered by Reconfigurable Intelligent Surfaces: How It Works, State of Research, and The Road Ahead," {\it IEEE Journal on Selected Areas in Communications}, vol. $38$, no. $11$, pp. $2450$\text{--}$2525$, $2020$. DOI: $10$.$1109$/JSAC.$2020$.$3007211$.

\bibitem[3]{Tang} W. Tang et al., "Wireless Communications With Reconfigurable Intelligent Surface: Path Loss Modeling and Experimental Measurement," {\it IEEE Transactions on Wireless Communications}, vol. $20$, no. $1$, pp. $421$\text{--}$439$, $2021$. DOI: $10$.$1109$/TWC.$2020$.$3024887$.

\bibitem[4]{Liu} Y. Liu et al., "Reconfigurable Intelligent Surfaces: Principles and Opportunities," {\it IEEE Communications Surveys \& Tutorials}, vol. $23$, no. $3$, pp. $1546$\text{--}$1577$, $2021$. DOI: $10$.$1109$/COMST.$2021$.$3077737$.

\bibitem[5]{SamorzewskiJTIT2023} A. Samorzewski, "Energy Consumption in Wireless Systems Equipped with RES, UAVs, and IRSs," {\it Journal of Telecommunications and Information Technology}, no. $2$, pp. $35\textbf{}${--}$40$, $2023$. DOI: $10$.$26636$/jtit.$2023$.$170923$.

\bibitem[6]{SamorzewskiKRiT2023} A. Samorzewski and A. Kliks "5G cellular systems supported by UAVs, RESs, and RISs," in {\it Radiocommunication and Teleinformatics Conference 2023 (pol. Konferencja Radiokomunikacji i Teleinformatyki 2023 -- KRiT 2023)}, Cracow, Poland, $2023$, pp.~$97\textbf{}${--}$100$. DOI: $10$.$15199$/$59$.$2023$.$4$.$18$.

\bibitem[7]{SamorzewskiSoftCOM2023} A. Samorzewski and A. Kliks, "5G Networks Supported by UAVs, RESs, and RISs," in {\it 2023 International Conference on Software, Telecommunications and Computer Networks (SoftCOM)}, Split, Croatia, $2023$, pp.~$1$\text{--}$6$. DOI: $10$.$23919$/SoftCOM$58365$.$2023$.$10271683$.

\bibitem[8]{SamorzewskiGLOBECOM2023} A. Samorzewski, M. Deruyck, and A. Kliks, "Energy Consumption in RES-Aware 5G Networks," in~{\it GLOBECOM 2023 -- 2023 IEEE Global Communications Conference}, Kuala Lumpur, Malaysia, $2023$, pp. $1024\textbf{}${--}$1029$. DOI: $10$.$1109$/GLOBECOM$54140$.$2023$.$10437451$.

\bibitem[9]{Castellanos} G. Castellanos, S. De Gheselle, L. Martens, N. Kuster, W. Joseph, M. Deruyck, and S. Kuehn, "Multi-objective optimization of human exposure for various $5$G network topologies in Switzerland," {\it Computer Networks}, vol. $2016$, $2022$. DOI: $10$.$1016$/j.comnet.$2022$.$109255$.

\bibitem[10]{NetworkData} {\it Database and map of BTS station locations / UKE permits}. BTSearch. [Online]. Available: http://beta.btsearch.pl.

\bibitem[11]{AreaData} {\it Poznan \text{--} Model 3D}. SIP (ang. Spatial Information System). [Online]. Available: http://sip.poznan.pl/model3d/\#/legend.

\bibitem[12]{3GPP} $3$GPP, "Technical Specification Group Radio Access Network; Study on channel model for frequencies from $0.5$ to $100$ GHz (Release $18$)," TR $38.901$ v$18.0.0$, $2024$.

\bibitem[13]{Bjornson} E. Björnson, J. Hoydis, and L. Sanguinetti, "Massive MIMO Networks: Spectral, Energy, and Hardware Efficiency," {\it Foundations and Trends® in Signal Processing}, vol. $11$, no. $3$\text{--}$4$, pp. $154$\text{--}$655$, $2017$. DOI: $10$.$1561$/$2000000093$.
\end{thebibliography}
\end{document}